\documentclass[preprint,prb,aps]{revtex4}
\usepackage{amsmath,amssymb}
\usepackage{graphicx}
\usepackage{dcolumn}
\usepackage{bm}

\begin{document}

\title[]{The Prelle-Singer method and Painlev\'e hierarchies}

\author{P~R~Gordoa$^a$, A~Pickering$^a$ and  M~Senthilvelan$^b$}
\date{16 May 2013, revised 2 April 2014}
\affiliation{$^a$ Departamento de Matem\'atica Aplicada,
Universidad Rey Juan Carlos, C/ Tulip\'an s/n, 28933 M\'ostoles,
Madrid, Spain,}
\affiliation{$^b$ Centre for Nonlinear Dynamics, School of Physics,
Bharathidasan University, Tiruchirapalli - 620 024, India.}

%\ead{$^{\dagger}$velan@cnld.bdu.ac.in}

\begin{abstract}

We consider systems of ordinary differential equations (ODEs) of the
form ${\cal B}{\mathbf K}=0$, where $\cal B$ is a Hamiltonian operator of
a completely integrable partial differential equation (PDE) hierarchy, and
${\mathbf K}=(K,L)^T$. Such systems, whilst of quite low order and linear
in the components of $\mathbf K$, may represent higher-order nonlinear
systems if we make a choice of $\mathbf K$ in terms of the coefficient
functions of $\cal B$. Indeed, our original motivation for the study of
such systems was their appearance in the study of Painlev\'e hierarchies,
where the question of the reduction of order  is of great importance.
However, here we do not consider such particular cases; instead we
study such systems for arbitrary $\mathbf K$, where they may represent
both integrable and nonintegrable systems of ordinary differential
equations. We consider the application of the Prelle-Singer (PS) method
--- a method used to find first integrals --- to such systems in order to
reduce their order. We consider the cases of coupled second order ODEs and
coupled third order ODEs, as well as the special case of a scalar third
order ODE; for the case of coupled third order ODEs, the development of the
PS method presented here is new. We apply the PS method to examples of such
systems, based on dispersive water wave, Ito and Korteweg-de Vries Hamiltonian
structures, and show that first integrals can be obtained. It is important to
remember that the equations in question may represent sequences of systems of
increasing order. We thus see that the PS method is a further technique
which we expect to be useful in our future work.

\end{abstract}


\maketitle

\section{Introduction}

In the last fifteen years or so, there has been a surge of interest
in Painlev\'e hierarchies, that is, hierarchies of ordinary differential
equations (ODEs) having the Painlev\'e property, as well as other
properties (underlying linear problems, B\"acklund transformations etc.)
possessed by the Painlev\'e equations. Such hierarchies are often derived
using associated hierarchies of completely integrable partial differential
equations (PDEs), either as similarity reductions \cite{A79,K97,GP11,GMP13}
or using non-isospectral extensions thereof. \cite{LRR92,GP99}

However, when one reduces a sequence of PDEs to a sequence of ODEs,
one is faced with the question of whether or not the order of
every member of that sequence of ODEs can be reduced. When this is
possible, we can then usually talk of a resulting Painlev\'e
hierarchy. For example, the dispersive water wave hierarchy results
in a sequence of ODEs the order of every member of which can be
reduced by two, and such that the first member of the sequence then
yields the fourth Painlev\'e equation: we thus talk of a fourth
Painlev\'e hierarchy. \cite{GJP01,GJP05,GJP06} From the dispersive
water wave hierarchy we can also derive a second Painlev\'e hierarchy
(a hierarchy of ODEs based on the second Painlev\'e equation).
\cite{GJP01,GJP06}

It is often the case that some of the equations resulting from the
reduction of an integrable PDE hierarchy are of the form
\begin{equation}
{\cal B}{\mathbf K}=0,
\label{bk}
\end{equation}
for some particular choice of $\mathbf K$, and where $\cal B$ is
one of the Hamiltonian operators of the PDE hierarchy. For example,
one case of reduction from the dispersive water wave hierarchy
gives (\ref{bk}) with \cite{P03,GJP05,GJP06,P12}
\begin{equation}
{\cal B}=\frac{1}{2}\left(\begin{array}{cc}
2\partial & \partial u -\partial^2 \\ u\partial +\partial^2 &
v\partial +\partial v
\end{array}\right)
\label{bb}
\end{equation}
where $\partial=\frac{\partial}{\partial x}$, $u=u(x)$ and $v=v(x)$,
for the choice
\begin{equation}
{\mathbf K}=\left(\begin{array}{c}K\\L\end{array}\right)=
{\bf L}_n+\sum_{i=1}^{n-1}c_i{\bf L}_i
+g_n\left(\begin{array}{c} 0 \\ x \end{array}\right).
\label{kk}
\end{equation}
Here $g_n$ and all $c_i$ are constants, and the quantities ${\bf L}_i$
(variational derivatives of a corresponding sequence of Hamiltonian
densities of the dispersive water wave hierarchy) are defined by the
recursion relation ${\cal C}{\bf L}_{i+1}={\cal B}{\bf L}_i$,
$i=0,1,2,\ldots$, where
\begin{equation}
{\cal C}=\left(\begin{array}{cc} 0 & \partial \\
      \partial & 0 \end{array}\right)
\end{equation}
is another of the three Hamiltonian operators of the dispersive
water wave hierarchy and ${\bf L}_0=(0,2)^T$. It is by reducing
the order of the system (\ref{bk}) that we then obtain a fourth
Painlev\'e hierarchy.

We considered the question of how to reduce the order of a system
which can be put in the form (\ref{bk}) in Refs.[\onlinecite{P03,GPP06}]. In
Refs.[\onlinecite{P03,GPP06}] we developed a method based on the factorization of
the Hamiltonian operator $\cal B$ under a Miura map. A discrete version
of this approach was given in Ref.[\onlinecite{GP09}]. In the present paper we
consider an alternative approach to the reduction of order of a system
(\ref{bk}), namely, the Prelle-Singer (PS) method. \cite{refs1,refs2,refs3}

The PS method has been successfully extended and used to derive first
integrals of a wide variety of systems of ODEs. \cite{refs4} However
here our emphasis is somewhat different. We are interested in systems
(\ref{bk}), where $\cal B$ is a Hamiltonian operator and $\mathbf K$
is left unspecified: we recall that the reductions of order we have
carried out thus far in order to obtain Painlev\'e hierarchies can in
fact be achieved independently of the form of $\mathbf K$ (the Painlev\'e
hierarchies then arise on specifying $\mathbf K$). Here we are therefore
concerned with systems (\ref{bk}), independently of whether they have a
connection with Painlev\'e hierarchies or not. Thus in the present
paper we are interested in systems (\ref{bk}) which represent a wide
variety of ODE systems, both integrable and nonintegrable (although
our original motivation was the integrable case). We then seek
first integrals of systems (\ref{bk}), treating them as systems in the
components of $\mathbf K$ with coefficients the functions appearing in
the Hamiltonian operator $\cal B$; these systems represent higher-order
systems in these coefficient functions once the components of
$\mathbf K$ have been specified in terms of these last, e.g., as in
(\ref{kk}). Thus we have an application of the reduction of order
of quite low-order systems (\ref{bk}), which are in fact linear in the
components of $\mathbf K$, to higher-order nonlinear systems.

The layout of the paper is as follows. In Section II we recall the PS
method for coupled second order ODEs. As an example we then consider
its application to the system (\ref{bk}) with $\cal B$ given by (\ref{bb}).
Thus we recover our previous results for the dispersive water wave case.
In Section III we extend the PS method to the case of coupled third order
ODEs; this represents a new extension of this method. As an application
we consider an example of Ito type. We also consider as a special case
of coupled third order ODEs an example of the case of a single third order
ODE. The final section is devoted to conclusions.

\section{\bf PS method for coupled second order ODEs}
\label{sec2}
Sometime ago Prelle and Singer [\onlinecite{GP0911}] have proposed a procedure for solving first order ODEs that presents the solution in terms of elementary functions if such a solution exists. The attractiveness of the PS method is that if the given system of first order ODEs has a solution in terms of elementary functions then the method guarantees that this solution will be found. Very recently Duarte et al [\onlinecite{GP0912}] have modified the technique developed by Prelle and Singer [\onlinecite{GP0911}] and applied it to second order ODEs. Their approach was based on the conjecture that if an elementary solution exists for the given second order ODE then there exists at least one elementary first integral $I(t, x, \dot{x})$ whose derivatives are all rational functions of $t$, $x$ and $\dot{x}$. ̇ For a class of systems these authors have deduced first integrals and in some cases for the first time through their procedure [\onlinecite{GP0912}]. Recently Chandrasekar et al have generalized the theory given in [\onlinecite{GP0912}] and pointed out a procedure to obtain all the integrals of motion/general solution and solved a class of nonlinear equations.\cite{refs1,refs2,refs3} Here we adopt the above said procedure for coupled second and third order ODEs.

Let us consider the following two coupled second order ODEs
\begin{eqnarray}
L_{xx}&=&\Phi_1(x,L,K,L_x,K_x),\nonumber\\
%L_{xx}&=& \;\;\;~~~~~~~~~~~~~~~~=
K_{xx}&=&\Phi_2(x,L,K,L_x,K_x).
\label{eL21}
\end{eqnarray}
Let us suppose that the system (\ref{eL21}) admits a
first integral of the form $I(x,L,K,L_{x},K_{x})=C$,
 with $C$ constant on the solutions so that the total differential gives
\begin{eqnarray}
dI={I_x}{dx}+{I_{L}}{dL}+{I_{K}}{dK}+{I_{L_{x}}}{dL_{x}}
+{I_{K_{x}}}{dK_{x}}=0. \label {cso02}
\end{eqnarray}
Here a subscript denotes partial differentiation with respect to the
corresponding variable and, by a common abuse of notation, we also
use subscripts (for example in equation (\ref{eL21}) where $K=K(x)$
and $L=L(x)$) to denote differentiation with respect to a unique
independent variable. Rewriting ({\ref{eL21}) in the form
\begin{eqnarray}
\Phi_1dx-dL_{x}=0,\qquad
\Phi_2dx-dK_{x}=0 \label {cso03}
\end{eqnarray}
and adding null terms
$s_1(x,L,K,L_{x},K_{x})L_{x}dx
- s_1(x,L,K,L_{x},K_{x})dL$ and $s_2(x,L,K,L_{x},K_{x})K_{x}dx
- s_2(x,L,K,L_{x},K_{x})dK$
 with the first equation in (\ref{cso03}), and
$u_1(x,L,K,L_{x},K_{x})L_{x}dx  -
u_1(x,L,K,L_{x},$ $K_{x})$ $dL $ and $u_2(x,L,K,L_{x},K_{x})K_{x}dx -
u_2(x,L,K,L_{x},K_{x})dK $ with the second equation in (\ref{cso03}),
respectively, we obtain that, on the solutions, the 1-forms
\begin{subequations}
\begin{eqnarray}
&&(\Phi_1+s_1L_{x}+s_2K_{x})dx-s_1dL-s_2dK-dL_{x}=0,\label {cso04}\\
&&(\Phi_2+u_1L_{x}+u_2K_{x})dx-u_1dL-u_2dK-dK_{x}=0.\label {cso05}
\end{eqnarray}
\label {cso06}
\end{subequations}

Hence, on the solutions, the 1-forms (\ref{cso02}) and
(\ref{cso06}) must be proportional. Multiplying (\ref{cso04}) by the factor
$ R_1(x,L,K,L_{x},K_{x})$ and (\ref{cso05}) by
the factor $ R_2(x,L,K,L_{x},K_{x})$, which act as the integrating
factors for (\ref{cso04}) and (\ref{cso05}), respectively, we have on the
solutions that
\begin{eqnarray}
dI=R_1(\Phi_1+SL_{x})dx+R_2(\Phi_2+UK_{x})dx-R_1SdL
-R_2UdK-R_1dL_{x}-R_2dK_{x}=0,\;\;\label {cso07}
\end{eqnarray}
where $S=\frac{R_1s_1+R_2u_1}{R_1}$ and $U=\frac{R_1s_2+R_2u_2}{R_2}$. Comparing
equations (\ref{cso07}) and (\ref{cso02}) we have, on the solutions,
the relations
\begin{eqnarray}
 I_x  =R_1(\Phi_1+SL_{x})+R_2(\Phi_2+UK_{x}),\; I_{L}  = -R_1S,
  I_{K} =-R_2U,\;
 I_{L_{x}}  =-R_1,\;
  I_{K_{x}} =-R_2.
 \nonumber \\ \label {cso08}
\end{eqnarray}
The compatibility conditions between equations (\ref{cso08}), namely $I_{xL}=I_{Lx},\;I_{xL_{x}}=I_{L_{x}x},\;I_{xK}=I_{Kx},\;I_{xK_{x}}=I_{K_{x}x},\;I_{LK}=I_{KL},\;I_{LL_{x}}=I_{L_{x}L},\;I_{KK_{x}}=I_{K_{x}K},\;I_{LK_{x}}=I_{K_{x}L},\;I_{KL_{x}}=I_{L_{x}K}$ and $I_{L_{x}K_{x}}=I_{K_{x}L_{x}}$, provide us with
the conditions,
\begin{eqnarray}
D[S] &=&-\Phi_{1L}-\frac{R_2}{R_1} \Phi_{2L}
    +\frac{R_2}{R_1}S\Phi_{2L_{x}}
           +S\Phi_{1L_{x}}+S^2,  \label {eq23}\\
D[U] &=&-\Phi_{2K}-\frac{R_1}{R_2}\Phi_{1K}
    +\frac{R_1}{R_2}U\Phi_{1K_{x}}
           +U\Phi_{2K_{x}}+U^2,  \label {eq24}\\
D[R_1]  &=&-(R_1\Phi_{1L_{x}}+R_2\Phi_{2L_{x}}+R_1 S),  \label {eq25}\\
D[R_2]  &=&-(R_2\Phi_{2K_{x}}+R_1\Phi_{1K_{x}}+R_2 U), \label {eq26}\\
SR_{1K} &=&-R_1S_{K}+UR_{2L}+R_2U_{L},\;\;R_{1L} =SR_{1L_{x}}+R_1S_{L_{x}},
  \label {eq31}\\
% \label {eq27}\\
R_{1K} &=&UR_{2L_{x}}+R_2U_{L_{x}},\qquad\qquad \;
%\label {eq28}\\
R_{2L} =SR_{1K_{x}}+R_1S_{K_{x}},  \label {eq29}\\
R_{2K}&=&UR_{2K_{x}}+R_2U_{K_{x}},\qquad\qquad\;
%\label {eq30}\\
R_{1K_{x}} =R_{2L_{x}}.\label {eq32}
\end{eqnarray}
Here the total differential operator, $D$, is defined by
$
D =\frac{\partial}{\partial{x}}+L_{x}\frac{\partial}{\partial{L}}
+K_{x}\frac{\partial}{\partial{K}}
+\Phi_1\frac{\partial}{\partial{L_{x}}}+\Phi_2\frac{\partial}
{\partial{K_{x}}}$. Integrating equations~(\ref{cso08}), we obtain the
integral of motion,
\begin{eqnarray}
I=r_1+r_2+r_3+r_4
-\int\bigg[R_2+\frac{\partial}{\partial K_{x}}\bigg(r_1+r_2+r_3+r_4\bigg)\bigg]
dK_{x},
\label{cso09}
\end{eqnarray}
where
\begin{eqnarray}
r_1&=\int\bigg(R_1(\Phi_1+SL_{x})+R_2(\Phi_2+UK_{x})\bigg)dx,
\qquad
r_2=-\int\bigg(R_1S+\frac{\partial}{\partial L}(r_1)\bigg)dL,&\qquad\nonumber\\
r_3&=-\int\bigg(R_2U+\frac{\partial}{\partial K}(r_1+r_2)\bigg)dK,\qquad
r_4=-\int\bigg[R_1+\frac{\partial}{\partial L_{x}}\bigg(r_1+r_2+r_3\bigg)\bigg]dL_{x}.&
\nonumber
\end{eqnarray}

Solving the determining equations,
(\ref{eq23})-(\ref{eq32}), consistently we can obtain expressions for the
functions $(S,U,R_1,R_2)$.
Substituting them into (\ref{cso09}) and evaluating the integrals we can construct
the associated integrals of motion. %The crux of the problem lies in solving the determining equations and identifying sufficient number of integrating factors and null forms. But the point is any particular solution of the given equation is an integrating factor for the given equation itself in the case of linear ODEs. As our motivation was to show this result.

%\newpage
\begin{center}
\bfseries Method of finding the integrating factors $R_1$ and $R_2$
\end{center}

To solve the determining equations for the functions
$S$, $U$, $R_1$ and $R_2$ we follow the following procedure.  Let us take a total derivative of equations (\ref{eq25}) and (\ref{eq26}). Doing so we get
\begin{align}
D^2[R_1]=-D[R_1\Phi_{1L_{x}}+R_2\Phi_{2L_{x}}+R_1S],\quad
%\label {3.1}\\
D^2[R_2]=-D[R_2\Phi_{2K_{x}}+R_1\Phi_{1K_{x}}+R_2U].
\label {3.1}
\end{align}
Using the identities
\begin{eqnarray}
D{[R_1 S]} =-(R_1\Phi_{1L}+R_2\Phi_{2L}),\quad D{[R_2U]} =-(R_1\Phi_{1K}+R_2\Phi_{2K}).
\label {eq33}
\end{eqnarray}
equation (\ref{3.1}) can be
rewritten in a coupled form for $R_1$ and $R_2$ as
\begin{align}
D^2[R_1]+D[R_1\Phi_{1L_{x}}+R_2\Phi_{2L_{x}}]=R_1\Phi_{1L}+R_2\Phi_{2L},
\label {tc01}\\
D^2[R_2]+D[R_2\Phi_{2K_{x}}+R_1\Phi_{1K_{x}}]=R_1\Phi_{1K}+R_2\Phi_{2K}.
\label {tc02}
\end{align}
The determining equations (\ref{tc01}) and (\ref{tc02}) form a system of linear PDEs in $R_1$ and $R_2$.  Substituting
the known expressions for $\Phi_1$ and $\Phi_2$ and their derivatives into (\ref{tc01}) and (\ref{tc02}) and solving them one can obtain expressions for the integrating factors $R_1$ and $R_2$. To obtain the explicit form of integrating factors $R_1$ and $R_2$ one may assume a suitable ansatz for these functions and substitute them into Eqs.(\ref{tc01}) and (\ref{tc02}) and solving the resultant equations consistently. Once $R_1$ and $R_2$ are known then the functions $(S,U)$ can be fixed through the relation (\ref{eq25})-(\ref{eq26}). Knowing $S,U,R_1$ and $R_2$, one has to make sure that the set $(S,U,R_1,R_2)$ also satisfies the remaining compatibility conditions (\ref{eq31}) - (\ref{eq32}). The set $(S,U,R_1,R_2)$ which satisfies all the equations (\ref{eq23}) - (\ref{eq32}) is then the acceptable solution and one can then determine the associated integral $I$ using the relation (\ref{cso09}). For complete integrability we require four independent compatible sets $(S_i,U_i,R_{1i},R_{2i}),\;i=1,2,3,4$. We note here that the examples which we consider in this paper are linear systems. In this case, as we show below, the determining equations (\ref{tc01}) and (\ref{tc02}) for the integrating factors coincide with the original system of equations (\ref{eL21}) which we consider initially. In other words any particular solution of the given equation forms an integrating factor for the given equation. So we do not discuss the method of solving the determining equations elaborately.
\begin{center}
\bfseries Example 1
\end{center}
%From (\ref{eL21}) we identify
Let us consider equation (\ref{bk}) with $\cal B$ given by (\ref{bb})
and $\mathbf K=(K,L)^T$, i.e.,
\begin{eqnarray}
L_{xx}&=&(u(x)L)_x+2K_x=\Phi_{1}, \nonumber \\
K_{xx}&=&-u(x)K_x-2v(x)L_x-Lv_x(x)=\Phi_{2}.
\label{eL23}
\end{eqnarray}
We recall that for the particular choice (\ref{kk}) this system (after reduction
of order by two) results in a fourth Painlev\'e hierarchy; we do not however,
make this choice here.
Substituting $\Phi_1, \Phi_2$ and their derivatives in (\ref{tc01}) and (\ref{tc02}) we get
\begin{eqnarray}
 D^2[R_1] & = & -D[u(x)R_1-2R_2v(x)]+R_1u_x(x)-R_2v_x(x),\nonumber\\
 D^2[R_2] & = & -D[-u(x)R_2+2R_1].
\label{eL24}
\end{eqnarray}
Rewriting
\begin{eqnarray}
 D^2[R_1]+D[u(x)R_1-2R_2v(x)]-R_1u_x(x)+R_2v_x(x) = 0,\nonumber\\
 D^2[R_2]+D[-u(x)R_2+2R_1] = 0.
\label{eL24a}
\end{eqnarray}
If we choose the integrating factors $R_2=-L$ and $R_1=K$ then Eq. (\ref{eL24a}) becomes
\begin{eqnarray}
 D^2[K]+D[u(x)K+2Lv(x)]-Ku_x(x)-Lv_x(x)=0,\nonumber\\
 -D^2[L]+D[u(x)L+2K]=0.
\label{eL24b}
\end{eqnarray}
The above equation is of the form
\begin{eqnarray}
K_{xx}+u(x)K_x+2v(x)L_x+Lv_x(x) & = 0,\nonumber\\
L_{xx}-(u(x)L)_x-2K_x & = 0,
\label{eL26}
\end{eqnarray}
which is nothing but (\ref{eL23}). Thus the integrating factors are
$R_1=K$ and $R_2=-L$.

Once $R_1$ and $R_2$ are known then the null forms $S$ and $U$ can be determined using
the ideas given in the theory.  They turn out to be
\begin{eqnarray}
S&=&-u(x)-2v(x)\frac{L}{K}-\frac{K_{x}}{K}, \nonumber\\
U&=& u(x)+2\frac{K}{L}-\frac{L_{x}}{L}.
\label{eL27}
\end{eqnarray}
Substituting the forms $R_2$, $R_1$, $S$ and $U$ into (\ref{cso09}) and evaluating the integrals
we find
\begin{eqnarray}
r_1 = LKu(x)+L^2v(x),\;\;
r_2 = LK_{x},\;\;
r_3 = K^2 - KL_{x}, \;\;
r_4 = 0
\label{eL28}
\end{eqnarray}
and
\begin{eqnarray}
I_1 = LKu(x)+L^2v(x)+LK_{x}+K^2-KL_{x}.
\label{eL29}
\end{eqnarray}
The first equation of (\ref{eL23}) can be integrated straightforwardly and yields the second integral
\begin{equation}
I_2=L_x-2K-u(x)L.
\end{equation}
The respective integrating factors are $R_1=1$ and $R_2=0$.
%%%%%%%%%%%%%%%%%%%%%%%%%%%%%%%%%%Sec2
\section{PS method for Coupled Third Order ODEs}
We now develop the PS method for coupled third order ODEs; this has not
appeared previously in the literature. We consider a system of the form
\begin{eqnarray}
L_{xxx}&=&\Phi_1(x,L,K,L_x,K_x,L_{xx},K_{xx}),\nonumber\\
K_{xxx}&=&\Phi_2(x,L,K,L_x,K_x,L_{xx},K_{xx}).
\label{vf1}
\end{eqnarray}
%\begin{eqnarray}
%K_{xxx}=-4(u_1K_x+\frac{1}{2}u_{1x}K+u_0L_x+\frac{1}{2}u_{0x}L-L_x)=\Phi_1(x,L,K,L_x,K_x,L_{xx},K_{xx}),\nonumber\\
%L_{xxx}=-4(u_1L_x+\frac{1}{2}u_{1x}L+u_0K_x+\frac{1}{2}u_{0x}K-K_x)=\Phi_2(x,L,K,L_x,K_x,L_{xx},K_{xx}).
%\label{vf1}
%%\end{eqnarray}
Let us assume that the above system admits a first integral of the form $I(x, L, K, L_x, K_x, L_{xx}, K_{xx})=C$, which is constant on the solutions. Then the total differentiation gives
\begin{equation}
dI=I_xdx+I_LdL+I_KdK+I_{L_x}dL_{x}+I_{K_x}dK_x+I_{L_{xx}}dL_{xx}+I_{K_{xx}}dK_{xx}=0.\label{int}
\end{equation}
We rewrite (\ref{vf1}) in the following form
\begin{equation}
\Phi_1dx-dL_{xx}=0,\;\;\;\;\Phi_2dx-dK_{xx}=0.
\end{equation}
Adding null terms in the above equation, finally we obtain the following equation
\begin{eqnarray}
(\Phi_1+S_1L_{x}+S_2K_x+M_1L_{xx}+M_2K_{xx})dx-S_1dL-S_2dK-M_1dL_{x}\nonumber \\-M_2dK_x-dL_{xx}=0,\label{e1}\\
(\Phi_2+U_1L_{x}+U_2K_x+N_1L_{xx}+N_2K_{xx})dx-U_1dL-U_2dK-N_1dL_{x}\nonumber\\-N_2dK_x-dK_{xx}=0.\label{e2}
\end{eqnarray}
Multiplying (\ref{e1}) by the integrating factor $R_1(x, L, K, L_x, K_x, L_{xx}, K_{xx})$ and (\ref{e2}) by \\$R_2(x, L, K, L_x, K_x, L_{xx}, K_{xx})$, we obtain the equation
\begin{eqnarray}
dI=R_1(\Phi_1+SL_{x}+ML_{xx})dx+R_2(\Phi_2+UK_x+NK_{xx})dx-R_1SdL\nonumber \\-R_2UdK-R_1MdL_{x}-R_2NdK_x-R_1dL_{xx}-R_2dK_{xx}=0
\end{eqnarray}
where $S=\frac{R_1S_1+R_2U_1} {R_1}$, $U=\frac{R_1S_2+R_2U_2} {R_2}$, $M=\frac{R_1M_1+R_2N_1} {R_1}$ and $N=\frac{R_1M_2+R_2N_2} {R_2}$.\\
Comparing the above equation with (\ref{int}), we obtain the following equations:
\begin{eqnarray}
I_x&=&R_1(\Phi_1+SL_{x}+ML_{xx})+R_2(\Phi_2+UK_x+NK_{xx}),\nonumber \\I_L&=&-R_1S,~~~~I_K=-R_2U,~~~~I_{L_x}=-R_1M,\nonumber \\
I_{K_x}&=&-R_2N,~~~~I_{L_{xx}}=-R_1,~~~~I_{K_{xx}}=-R_2.\label{int1}
\end{eqnarray}
Using the compatibility conditions, we obtain the following determining equations
\begin{eqnarray}
D[R_1]&=&-(R_1\Phi_{1L_{xx}}+R_2\Phi_{2L_{xx}}+R_1M),\label{d1}\\
D[R_2]&=&-(R_1\Phi_{1K_{xx}}+R_2\Phi_{2K_{xx}}+R_2N),\label{d2}\\
D[S]&=&S\Phi_{1L_{xx}}+(\frac{SR_2} {R_1})\Phi_{2L_{xx}}+MS-\Phi_{1L}-(\frac{R_2} {R_1})\Phi_{2L},\label{d3}\\
D[U]&=&U\Phi_{2K_{xx}}+(\frac{R_1U} {R_2})\Phi_{1K_{xx}}+NU-\Phi_{2K}-(\frac{R_1} {R_2})\Phi_{1K},\label{d4}\\
D[M]&=&M\Phi_{1L_{xx}}+(\frac{MR_2} {R_1})\Phi_{2L_{xx}}+M^2-\Phi_{1L_{x}}-(\frac{R_2} {R_1})\Phi_{2L_{x}}-S,\label{d5}\\
D[N]&=&N\Phi_{2K_{xx}}+(\frac{R_1N} {R_2})\Phi_{1K_{xx}}+N^2-\Phi_{2K_x}-(\frac{R_1} {R_2})\Phi_{1K_x}-U,\label{d6}\\
R_{1K_x}M+R_1M_{K_x}&=&R_{2L_x}N+R_2N_{L_x},\;\;\;\;\;R_{1L}=R_{1L_{xx}}S+R_1S_{L_{xx}},\label{d7}\\
R_{1K_x}S+R_1S_{K_x}&=&R_{2L}N+R_2N_L,\;\;\;\;\;R_{1K}=R_{2L_{xx}}U+R_2U_{L_{xx}},\label{d8}\\
R_{2K_x}U+R_2U_{K_x}&=&R_{2K}N+R_2N_K,\;\;\;\;\;R_{1L_x}=R_{1L_{xx}}M+R_1M_{L_{xx}},\label{d9}\\
R_{1L_x}S+R_1S_{L_x}&=&R_{1L}M+R_1M_L,\;\;\;\;\;R_{1K_x}=R_{2L_{xx}}N+R_2N_{L_{xx}},\label{d10}\\
R_{2L_x}U+R_2U_{L_x}&=&R_1M_K+MR_{1K},\;\;\;\;\;R_{2L}=R_{1K_{xx}}S+R_1S_{K_{xx}},\label{d11}\\
R_{1K}S+R_1S_K&=&R_{2L}U+R_2U_L,\;\;\;\;\;R_{2K}=R_{2K_{xx}}U+R_2U_{K_{xx}},\label{d12}\\
R_{2L_x}&=&R_{1K_{xx}}M+R_1M_{K_{xx}},\;\;\;\;\;R_{2K_x}=R_{2K_{xx}}N+R_2N_{K_{xx}},\label{d13}\\
R_{1K_{xx}}&=&R_{2L_{xx}}\label{d14}.
\end{eqnarray}
Once the compatible solution, $R_1, R_2, S, U, M$ and $N$ are determined then the integral can be readily constructed by substituting all the expressions in (\ref{int1}) and integrating the resulting equation, that is
\begin{equation}
I=r_1+r_2+r_3+r_4+r_5+r_6-\int(R_2+\frac{\partial} {\partial K_{xx}}(r_1+r_2+r_3+r_4+r_5+r_6))dK_{xx}
\label{lkl}
\end{equation}
where
\begin{eqnarray}
r_1&=&\int[R_1(\Phi_1+SL_{x}+ML_{xx})+R_2(\Phi_2+UK_x+NK_{xx})]dx,\nonumber \\
r_2&=&-\int(R_1S+\frac{\partial} {\partial L}(r_1))dL,\nonumber \\
r_3&=&-\int(R_2U+\frac{\partial} {\partial K}(r_1+r_2))dK,\nonumber \\
r_4&=&-\int(R_1M+\frac{\partial} {\partial L_{x}}(r_1+r_2+r_3))dL_{x},\nonumber
\end{eqnarray}
\begin{eqnarray}
r_5&=&-\int(R_2N+\frac{\partial} {\partial K_x}(r_1+r_2+r_3+r_4))dK_x,\nonumber \\
r_6&=&-\int(R_1+\frac{\partial} {\partial L_{xx}}(r_1+r_2+r_3+r_4+r_5))dL_{xx}.\nonumber
\end{eqnarray}
As we did in the second order case to determine the integrating factors $R_1$ and $R_2$ we rewrite the determining equations (\ref{d1})-(\ref{d6}) as two equations.
\begin{eqnarray}
D^3[R_1]+D^2[R_1\Phi_{1L_{xx}}+R_2\Phi_{2L_{xx}}]-D[R_1\Phi_{1L_{x}}+R_2\Phi_{2L_{x}}]+R_1\Phi_{1L}+R_2\Phi_{2L}=0,\label{dd1}\\
D^3[R_2]+D^2[R_1\Phi_{1K_{xx}}+R_2\Phi_{2K_{xx}}]-D[R_1\Phi_{1K_x}+R_2\Phi_{2K_x}]+R_1\Phi_{1K}+R_2\Phi_{2K}=0,
\label{dd2}
\end{eqnarray}
where $D=\frac{\partial} {\partial x}+L_x\frac{\partial} {\partial L}+K_x\frac{\partial} {\partial K}+L_{xx}\frac{\partial} {\partial L_x}+K_{xx}\frac{\partial} {\partial K_x}+\Phi_1\frac{\partial} {\partial L_{xx}}+\Phi_2\frac{\partial} {\partial K_{xx}}$.
%Once we find $R_1$ and $R_2$ then we can derive all other quantities from using the equations (\ref{d3})-(\ref{d6}). Once all the functions are known, we can construct the integral using (\ref{lkl}).

The determining equations (\ref{dd1}) and
(\ref{dd2}) form a system of linear PDEs in $R_1$ and $R_2$. Substituting
the known expressions for $\Phi_1$ and $\Phi_2$ and their derivatives
into (\ref{dd1}) and
(\ref{dd2}) and solving them one can obtain expressions for the
integrating factors $R_1$ and $R_2$. Once $R_1$ and $R_2$ are known then the
functions $(S,U,M,N)$ can be fixed through the relations
(\ref{d3})-(\ref{d6}). Knowing $S,U,M,N,R_1$ and $R_2$, one has to make
sure that the set $(S,U,M,N,R_1,R_2)$ also satisfies the remaining
compatibility conditions (\ref{d7}) - (\ref{d14}). The set
$(S,U,M,N,R_1,R_2)$ which satisfies all the equations (\ref{d1}) -
(\ref{d14}) is then the acceptable solution and one can then
determine the associated integral $I$ using the relation
(\ref{lkl}). \\
If we choose $K=0$ and $\Phi_2=0$ in the above procedure, we get the Prelle-Singer procedure for scalar third order ODEs. We also present an example for this case.
\begin{center}
\bfseries Example 2
\end{center}
Let us consider a system of Ito type, i.e., of the form (\ref{bk}) with
\cite{lm91}
\begin{equation}
{\cal B}=\left(\begin{array}{cc}
\frac{1}{4}\partial^3+u_1\partial+\frac{1}{2}u_{1x} &
u_0\partial+\frac{1}{2}u_{0x}-\partial \\
u_0\partial+\frac{1}{2}u_{0x}-\partial &
\frac{1}{4}\partial^3+u_1\partial+\frac{1}{2}u_{1x}
\end{array}\right),
\end{equation}
where $u_0=u_0(x)$, $u_1=u_1(x)$ and ${\mathbf K}=(K,L)^T$. This system we
write as
%From (\ref{vf1}) we identify
\begin{eqnarray}
\displaystyle L_{xxx}=-4(u_1L_x+\frac{1}{2}u_{1x}L+u_0K_x+\frac{1}{2}u_{0x}K-K_x)=\Phi_1(x,L,K,L_x,K_x,L_{xx},K_{xx}),\nonumber\\
\displaystyle K_{xxx}=-4(u_1K_x+\frac{1}{2}u_{1x}K+u_0L_x+\frac{1}{2}u_{0x}L-L_x)=\Phi_2(x,L,K,L_x,K_x,L_{xx},K_{xx}).
\label{vgf1}
\end{eqnarray}
We are interested in showing that we can use the PS method to construct first
integrals of this system in the general case, i.e., independently
of whether for a particular choice of $\mathbf K$ the system is integrable or
nonintegrable, or of whether there is a connection with Painlev\'e hierarchies.
%\begin{eqnarray}
%\Phi_{1L}&=&-2u_{1x}(x),\;\;\;\;\;~~~~~\Phi_{1K}=-2u_{0x}(x),\nonumber\\
%\Phi_{1L_{x}} &=&-4u_1(x),\;\;\;\;\;~~~~~\Phi_{1K_x}=-4(u_0(x)-1),\nonumber\\
%\Phi_{1L_{xx}}&=&0,\;\;\;\;\;\;\;~~~~~~~~~~~\Phi_{1K_{xx}}=0 \nonumber\\
%\Phi_{2L}&=&-2u_{0x}(x),\;\;\;\;\;~~~~~\Phi_{2K}=-2u_{1x}(x),\nonumber\\
%\Phi_{2L_{x}} & =&-4(u_0(x)-1),\;\;\Phi_{2K_x}=-4u_1(x), \nonumber\\
%\Phi_{2L_{xx}}&=&0, \;\;\;\;\;\;\;~~~~~~~~~~~\Phi_{2K_{xx}}=0
%\label{eL23}
%\end{eqnarray}
Substituting $\Phi_1, \Phi_2$ and their derivatives in (\ref{dd1}) and (\ref{dd2}) we find
\begin{eqnarray}
\displaystyle D^3[R_1]-D[R_1(-4u_1(x))+R_2(-4u_0(x)+4)]+R_1(-2u_{1x}(x))+R_2(-2u_{0x}(x))=0,\nonumber \\
\displaystyle D^3[R_2]-D[R_1(-4u_0(x)+4)+R_2(-4u_1(x))]+R_1(-2u_{0x}(x))+R_2(-2u_{1x}(x))=0.\label{f1}
\end{eqnarray}
As we noted earlier for linear ODEs,  the determining equations for the integrating factors coincide with the original ODEs. If we choose the integrating factors $R_2=L$ and $R_1=K$ then Eq.(\ref{f1}) exactly coincides with (\ref{vgf1}).
%\begin{eqnarray}
%D^3[K]-D[K(-4u_1(x))+L(-4u_0(x)+4)]+K(-2u_{1x}(x))+L(-2u_{0x})&=0,\nonumber \\
%D^3[L]-D[K(-4u_0(x)+4)+L(-4u_1(x))]+K(-2u_{0x}(x))+L(-2u_{1x})&=0.\label{fr}
%\end{eqnarray}
%The above equation is nothing but (\ref{vf1}).
Thus the integrating factors are $R_2=L$ and $R_1=K$.
Once $R_1$ and $R_2$ are known the null forms $S,U,M$ and $N$ can be determined using the ideas given in the previous section. They turn out to be
\begin{eqnarray}
M&=&-\frac{K_x} {K},\;\;\;S=4u_1(x)+\frac{4Lu_0(x)-4L+K_{xx}} {K},\nonumber \\
N&=&-\frac{L_x} {L},~~\;\;\;U=4u_1(x)+\frac{4Ku_0(x)-4K+L_{xx}} {L}.
\end{eqnarray}
Substituting the above equations in (\ref{lkl}) we find
\begin{eqnarray}
r_1&=&-4LKu_{1}(x)-2u_{0}(x)(L^2+K^2),\;\;\;r_2=2L^2-K_{xx}L,\nonumber \\
r_3&=&2K^2-L_{xx}K,\;\;\;r_4=L_xK_x,\;\;r_5=0,\;\;r_6=0.
\end{eqnarray}
and the first integral is of the form
\begin{equation}
I_1=-4\left(LKu_1(x)+\frac{1} {2}u_0(x)(L^2+K^2)-\frac{1} {2}L^2+\frac{1} {4}K_{xx}L-\frac{1} {2}K^2+\frac{1} {4}L_{xx}K-\frac{1} {4}L_xK_x\right).
\end{equation}
The second set of integrating factors can also be easily fixed from Eq.(\ref{f1}) as $R_2=K$ and $R_1=L$. The associated null terms turn out to be
\begin{eqnarray}
M&=&-\frac{L_x} {L},\;\;S=4u_1(x)+\frac{4Ku_0(x)-4K+L_{xx}} {L},\nonumber \\
N&=&-\frac{K_x} {K},~~\;\;U=4u_1(x)+\frac{4Lu_0(x)-4L+K_{xx}} {K}.
\end{eqnarray}
Substituting the above equations in (\ref{lkl}) we find
\begin{eqnarray}
r_1&=&-4LKu_0(x)-2u_1(x)(L^2+K^2),\;\;\;r_2=4KL-L_{xx}L,\nonumber \\
r_3&=&-K_{xx}K,\;\;\;\;\;\;r_4=\frac{L_{x}^2} {2},\;\;r_5=\frac{K_{x}^2} {2},\;\;\;r_6=0.
\end{eqnarray}
The second integral is found to be of the form
\begin{equation}
I_2=-4\left(\frac{1}{2}u_1(x)(L^2+K^2)+u_0(x)KL-KL+\frac{1} {4}L_{xx}L+\frac{1} {4}K_{xx}K-\frac{1} {8}L_{x}^2-\frac{1} {8}K_{x}^2\right).
\end{equation}
Whilst this example was considered in Ref.[\onlinecite{GPP06}] using a technique
based on the factorization of the Hamiltonian operator $\cal B$ under a Miura
map, first integrals of the system (\ref{vgf1}) were not explicitly given
in this general case (first integrals were given for a particular case, i.e.,
for a particular choice of $\mathbf K$).
\begin{center}
\bfseries Example 3
\end{center}
Let us consider the equation
\begin{equation}
{\cal B}L=[\partial^3+4u\partial+2u_x]L=0
\end{equation}
where $u=u(x)$, i.e.,
\begin{eqnarray}
L_{xxx}=\Phi_1 = -4u(x)L_{x}-2u_x(x)L.
%\;\; \Phi_{L_{x}} =  -4u(x), \;\;\Phi_L = -2u_x(x).
\label{eL3}
\end{eqnarray}
For a particular choice of $L$, this equation leads us (after integration)
to a thirty-fourth Painlev\'e hierarchy. \cite{AH98,CJP99} Once again,
we do not make this choice of $L$ here, as we are interested in both
integrable and nonintegrable cases. Substituting Eq. (\ref{eL3}) into
(\ref{dd1}) we get
\begin{eqnarray}
 D^3[R_1]+4D[u(x)R_1]-2u_x(x)R_1=0,
\label{eL4}
\end{eqnarray}
which can be written
\begin{eqnarray}
 D^3[R_1]+4u(x)D[R_1]+2u_x(x)R_1=0.
\label{eL5}
\end{eqnarray}
%This is of the form
%\begin{eqnarray}
%R_{1xxx}+4u(x)R_{1x}+2u_x(x)R_1=0 \,\,
%\label{eL5a}
%\end{eqnarray}
Choosing $R_1=L$, we see that equation (\ref{eL5}) exactly coincides with
(\ref{eL3}).  As a consequence
one simple solution (integrating factor) of (\ref{eL3}) can be immediately written in the form
\begin{eqnarray}
 R_1 = L.
\label{eL6}
\end{eqnarray}
In other words one integrating factor for the
equation (\ref{eL3}) is $L$ itself.
Once $R_1$ is known then $M$ can be determined straightforwardly from Eq. (\ref{d1}).  Simplifying the resulting expression we find
\begin{eqnarray}
M = -\frac{L_{x}}{L}.
\label{eL7}
\end{eqnarray}
Substituting $M$ into (\ref{d3}) we find that $S$ is of the form
\begin{eqnarray}
S = 4u(x)+\frac{L_{xx}}{L}.
\label{eL8}
\end{eqnarray}
Plugging the expressions $R_1$, $S$ and $M$ into the expression (\ref{lkl}) and evaluating the integrals
we find
\begin{equation}
r_1=-2L^2u(x),\;\; r_2 = -LL_{xx},\;\; r_4 = \frac{L_x^2}{2},\;\; r_3=r_5=r_6=0,\nonumber
\end{equation}
and
\begin{equation}
 I = -\left(LL_{xx}+2L^2u(x)-\displaystyle{\frac{L_{x}^2}{2}}\right).\label{eL9}
\end{equation}
Of course, in this example, the integrating factor can be identified by
inspection; our aim here is to show that the PS method is also applicable in
the scalar case of third order equations which may in fact, for particular
choices of $L$, represent higher order ODEs.
One can easily check that $\frac{dI} {dx}=0$.

\section{Conclusions}

We have considered the application of the PS method to systems of the form
(\ref{bk}), where $\cal B$ is a Hamiltonian operator of a completely
integrable PDE hierarchy, and ${\mathbf K}=(K,L)^T$. The original motivation
for the study of such systems was their appearance in the study of Painlev\'e
hierarchies. However, we are also interested in the study of such systems
outside of that context, where they may also represent nonintegrable systems.
It is interesting that these quite low-order systems, linear in the components
of  $\mathbf K$, may represent higher-order nonlinear systems.
We have considered the cases of coupled second order ODEs and coupled third
order ODEs, as well as the special case of a scalar third order ODE; for the
case of coupled third order ODEs, the development of the PS method presented
here is new. We have  successfully applied the PS method to examples of such
systems, and have succeeded in obtaining first integrals. This then represents
a further technique, additional to the factorization of $\cal B$ under a Miura
map, which we expect to be of use in our future work.

\section{Acknowledgements}
The work of PRG and AP was supported in part by the Ministry of Science
and Innovation of Spain under contract MTM2009-12670. The work of PRG and AP
is currently supported by the Ministry of Economy and Competitiveness
of Spain under contract MTM2012-37070. This work was undertaken in part
during a visit by MS to the Universidad Rey Juan Carlos, Madrid, in November
2012, financed by the project MTM2009-12670.

\end{document}